# Measurement, interpretation and information

Olimpia Lombardi [1,†,*], Sebastian Fortin [2,†] and Cristian López [3,†]

[1] CONICET and University of Buenos Aires; E-Mail: olimpiafilo@arnet.com.ar
[2] CONICET and University of Buenos Aires; E-Mail: sfortin@gmx.net
[3] CONICET and University of Buenos Aires; E-Mail: lopez.cristian1987@gmail.com

[†] These authors contributed equally to this work.

**Abstract:** During many years since the birth of quantum mechanics, instrumentalist interpretations prevailed: the meaning of the theory was expressed in terms of measurements results. But in the last decades, several attempts to interpret it from a realist viewpoint have been proposed. Among them, modal interpretations supply a realist non-collapse account, according to which the system always has definite properties and the quantum state represents possibilities, not actualities. However, the traditional modal interpretations faced some conceptual problems when addressing imperfect measurements. The modal-Hamiltonian interpretation, on the contrary, proved to be able to supply an adequate account of the measurement problem, both in its ideal and its non-ideal versions. Moreover, in the non-ideal case, it gives a precise criterion to distinguish between reliable and non-reliable measurements. Nevertheless, that criterion depends on the particular state of the measured system, and this might be considered as a shortcoming of the proposal. In fact, one could ask for a criterion of reliability that does not depend on the features of what is measured but only on the properties of the measurement device. The aim of this article is precisely to supply such a criterion: we will adopt an informational perspective for this purpose.

## 1. Introduction

In classical physics, it is supposed that measurement is a topic of interest exclusively for experimental physicists, and that theories can be studied without considering how the information about the system will be obtained. But this is not true: in classical statistical mechanics, measurement affects entropy and is relevant with respect to the possibility of Maxwell's demon (see [1]). Quantum mechanics, in turn, places measurement in the center of the stage: the acquisition of information about the measured system turns out to be a theoretical problem in itself. In fact, the attempt to solve the measurement problem has traditionally been the main motivation for most interpretations of quantum mechanics.

During many years since the birth of the theory, instrumentalist interpretations prevailed: the meaning of quantum mechanics was expressed in terms of measurements results. But in the last decades, several attempts to interpret quantum mechanics from a realist viewpoint have been proposed. Among them, modal interpretations (see [2]) supply a realist non-collapse account, according to which the system has definite properties at all times and the quantum state represents possibilities, not actualities. However, the traditional modal interpretations faced some conceptual problems when addressing imperfect measurements. The modal-Hamiltonian interpretation, on the contrary, proved to be able to supply an adequate account of the measurement problem, both in its ideal and its non-ideal versions.

In this sense, an advantage of the modal-Hamiltonian interpretation is that, in the non-ideal case, it gives a precise criterion to distinguish between reliable and non-reliable measurements. Nevertheless, that criterion depends on the particular state of the measured system, and this might be considered as a shortcoming of the proposal. In fact, one could ask for a criterion of reliability that does not depend on the features of what is measured but only on the properties of the measurement device. The aim of this article is precisely to supply such a criterion, and we will adopt an informational perspective for this purpose.

In order to fulfill this task, the article is organized as follows. In Section 2, the measurement problem will be formulated with precision, both in its ideal and its non-ideal versions. In Section 3, the modal interpretations of quantum mechanics are introduced, focusing in particular in the modal-Hamiltonian interpretation. Section 4 will be entirely devoted to show how the modal-Hamiltonian interpretation solves the measurement problem, and to explain the criterion of reliable measurement proposed for the non-ideal case. In Section 5, we will reconstruct measurement from an informational perspective, which will allow us to reformulate the reliability criterion in a conceptually more adequate way. Finally, In Section 6 we will introduce our conclusions, based on the idea that a quantum measurement can be viewed as an informational process precisely characterized in terms of the Shannon theory.

**2. The quantum measurement problem**

*2.1. Three concepts of quantum measurement*

In the standard von Neumann model, a quantum measurement is conceived as an interaction between a measured system $S$ and a measuring apparatus $M$. Before the interaction, $M$ is prepared in a ready-to-measure state $|p_0\rangle$, eigenvector of the pointer observable $P$ of $M$, and the state of $S$ is a superposition of the eigenstates $|a_i\rangle$ of an observable $A$ of $S$. The interaction introduces a correlation between the eigenstates $|a_i\rangle$ of $A$ and the eigenstates $|p_i\rangle$ of $P$:

$$|\psi_0\rangle = \sum_i \alpha_i |a_i\rangle \otimes |p_0\rangle \quad \rightarrow \quad |\psi\rangle = \sum_i \alpha_i |a_i\rangle \otimes |p_i\rangle \tag{1}$$

The problem consists in explaining why, being the state $|\psi\rangle$ a superposition of the $|a_i\rangle \otimes |p_i\rangle$, the pointer $P$ acquires a definite value.

In the orthodox collapse interpretation, the pure state $|\psi\rangle$ is assumed to "collapse" to one of the components of the superposition, say $|a_k\rangle \otimes |p_k\rangle$, with probability $|\alpha_k|^2$:

$$|\psi_0\rangle = \sum_i \alpha_i |a_i\rangle \otimes |p_0\rangle \xrightarrow{collapse} |a_k\rangle \otimes |p_k\rangle \tag{2}$$

In this situation it is supposed that the measuring apparatus is in one of the eigenstates $|p_i\rangle$ of $P$, in this case $|p_k\rangle$, and therefore $P$ acquires a definite value $p_k$, the eigenvalue corresponding to the eigenstate $|p_k\rangle$, with probability $|\alpha_k|^2$. As a consequence, the state of the composite system $S+M$ after measurement is represented by a mixture $\rho^c$:

$$\rho^c = \sum_i |\alpha_i|^2 |a_i\rangle \otimes |p_i\rangle \langle a_i| \otimes \langle p_i| \tag{3}$$

where the probabilities $|\alpha_i|^2$ are endowed with an ignorance interpretation.

This von Neumann model of quantum measurement could be easily interpreted under the paradigm of classical measurements, which are based on the correlation between the actual values of an apparatus' pointer and of an observable to be measured. But this reading does not take into account the difference between classical and quantum measurements. In classical measurement, the state of the system is measured by revealing the value of the observables that define that state. In quantum mechanics, by contrast, the value of an observable can be determined by measurement only when the system's state is an eigenstate of that observable; in other cases, the aim is to reconstruct the state of the system just before the beginning of the measurement process. For this purpose, it is necessary to know the coefficients of the state in different bases; such coefficients are inferred from probabilities. This means that, due to the probabilistic nature of quantum mechanics, in general the maximum information about a quantum system is obtained by means of repeated measurements on the same system or on identical systems. On this basis, three different concepts of quantum measurement should be distinguished:

- *Single measurement*: It is a single process, in which the reading of the pointer is registered. A single measurement, considered in isolation, does not supply yet relevant information about the state of the system $S$ prior to the measurement, since the amplitudes are not revealed.

- *Frequency measurement*: It is a repetition of identical single measurements, whose purpose is to obtain the values $|\alpha_i|^2$ on the basis of the frequencies of the pointer readings in the different single measurements. A frequency measurement supplies relevant information about the state of $S$, but is not yet sufficient to completely identify such a state, since it does not reveal the phases.

- *State measurement*: It is a collection of frequency measurements, each one of them with its particular experimental arrangement. Each arrangement correlates the pointer $P$ of the apparatus $M$ with an observable $A_i$ of the system, in such a way that the $A_i$ are not only different, but also non-commuting to each other. The information obtained by means of such a collection of frequency measurements is sufficient to reconstruct the state of $S$ (see [3]).

If the quantum state is conceived as the state of a single system (and not as referring to an ensemble, as in the so-called "ensemble" or "statistical" interpretation" proposed by Leslie Ballentine [4]), the von Neumann scheme, understood as a model for measurement, addresses single measurements. This is completely reasonable to the extent that, if we do not have an adequate explanation of the single case, we cannot account for the results obtained by the repetition of single cases. Nevertheless, it should not be forgotten that a single measurement is always an element of a measurement procedure by means of which, finally, frequencies are to be obtained.

*2.2. Ideal and non ideal measurements*

As it is well known, the ideal measurement is a very special and idealized case. For this reason, it is necessary to consider non-ideal situations. Two kinds of non-ideal measurements are usually distinguished in the literature:

- *Imperfect measurement* (*first kind*):

$$\sum_i \alpha_i |a_i\rangle \otimes |p_0\rangle \;\rightarrow\; \sum_{ij} \beta_{ij} |a_i\rangle \otimes |p_j\rangle \tag{4}$$

where, in general, $\beta_{ij} \neq 0$ with $i \neq j$.

- *Disturbing measurement* (*second kind*):

$$\sum_i \alpha_i |a_i\rangle \otimes |p_0\rangle \;\rightarrow\; \sum_i \alpha_i |a_i^d\rangle \otimes |p_i\rangle \qquad (5)$$

where, in general, $\langle a_i^d | a_j^d \rangle \neq \delta_{ij}$

In the case of the *imperfect measurement*, the interaction Hamiltonian fails to establish a perfect correlation between the eigenstates $|a_i\rangle$ of the observable $A$ of the measured system $S$ and the eigenstates $|p_i\rangle$ of the pointer $P$ of the measuring apparatus $M$ (see eq. (4)). But, in this case, the measurement does not disturb the original state of the measured system, that is, the eigenstates $|a_i\rangle$ are not modified by the interaction. Nevertheless, if the interaction were to still count as a measurement in spite of imperfection, the $\beta_{ij} \neq 0$, with $i \neq j$, should be small with respect to the $\beta_{ii}$.

In the case of the *disturbing measurement*, by contrast, although the interaction does not introduce non-diagonal terms, it modifies the state of the original system. In particular, each eigenstate $|a_i\rangle$ becomes an $|a_i^d\rangle \neq |a_i\rangle$, and the $|a_i^d\rangle$ do not need to be orthogonal to each other (see eq. (5)). Nevertheless, if the interaction were to still count as a measurement in spite of disturbance, the $|a_i^d\rangle$ should not be very different than the corresponding original $|a_i\rangle$.

Notwithstanding the difference between the two sources of non-ideality, the disturbing measurement can be expressed under the form of an imperfect measurement by a change of basis:

$$\sum_i \alpha_i |a_i^d\rangle \otimes |p_i\rangle = \sum_{ij} \beta_{ij} |a_i\rangle \otimes |p_j\rangle \qquad (6)$$

Then, in general we will talk about non-ideal measurements without distinction (however, the difference will be relevant to the traditional modal interpretations, see next section).

Of course, ideal measurement is a situation that can never be achieved in practice: the interaction between the measured system and the measuring apparatus never introduces a completely perfect correlation nor leaves the initial state absolutely undisturbed. In spite of this, physicists usually perform measurements pragmatically considered successful. Therefore, an interpretation of quantum mechanics should be capable of accounting for both ideal and non-ideal measurements. As we will see, this will be the Achilles heel of certain interpretations specifically designed to solve the measurement problem.

### 3. Modal interpretations of quantum mechanics

*3.1. The modal family*

Modal interpretations find their roots in the works of Bas van Fraassen ([5], [6], [7]), where the distinction between the quantum state and −what he called− the *value state* of the system is introduced: the quantum state tells us what *may* be the case, that is, which physical properties the system may possess; the value state represents what *actually* is the case. Therefore, the quantum state is the basis for modal statements, that is, statements about what possibly or necessarily is the case.

On the basis of this original idea, in the decade of 1980 several authors developed realist interpretations that, retrospectively, can be regarded as new elaborations on van Fraassen's proposal (for an overview and references, see [2]). The members of this family of interpretations share the following features:

- The interpretation is based on the standard formalism of quantum mechanics.

- The interpretation is realist, that is, it aims at describing how reality would be if quantum mechanics were true.
- Quantum mechanics is a fundamental theory: it describes not only elementary particles but also macroscopic objects; quantum states refer to single systems, not to ensembles.
- The quantum state describes the probabilities of the possible properties of the system. In turn, systems possess actual properties at all times, whether or not a measurement is performed on them. The relationship between the quantum state and the actual properties is probabilistic.
- A quantum measurement is an ordinary physical interaction. There is no collapse: the quantum state always evolves unitarily according to the Schrödinger equation, which gives the time evolution of probabilities, not of actual properties.

The *contextuality* of quantum mechanics, expressed by the Kochen-Specker theorem ([8]), is a barrier to any realist classical-like interpretation of QM, since it proves the impossibility of ascribing precise values to all the physical properties (observables) of a quantum system simultaneously. Therefore, realist non-collapse interpretations are committed to selecting a privileged set of definite-valued observables. Each modal interpretation supplies a "rule of definite-value ascription" or "actualization rule", which picks out, from the set of all observables of a quantum system, the subset of definite-valued properties that constitute the preferred context.

The traditional Kochen-Dieks modal interpretation ([9]-[15]) is based on the biorthogonal (Schmidt) decomposition of the pure quantum state of the system, according to which the state picks out (in many cases, uniquely) a basis for each of the subsystems. According to this interpretation, those bases generate the definite-valued properties of the corresponding subsystems. This traditional view is particularly appropriate to account for quantum measurement. In fact, given the correlation introduced by the interaction between the measured system $S$ and the measuring apparatus $M$ (see eq. (1)), the preferred context of $S$ is defined by the set $\{|a_i\rangle\}$ and the preferred context of $M$ is defined by the set $\{|p_i\rangle\}$. Therefore, the pointer position $P$ is a definite-valued property of the apparatus: it acquires one of its possible values (eigenvalues) $p_i$. And analogously in the measured system: the measured observable $A$ is a definite-valued property of the measured system, and it acquires one of its possible values (eigenvalues) $a_i$.

The Vermaas-Dieks modal interpretation ([16]) is a generalization of the previous one to mixed states, based on the spectral decomposition of the reduced density operator: the definite-valued properties $\Pi_i$ of a system and their corresponding probabilities $\Pr_i$ are given by the non-zero diagonal elements of the spectral decomposition of the system's state.

$$\rho = \sum_i \gamma_i \Pi_i \qquad \Pr_i = Tr(\rho \Pi_i) \qquad (7)$$

This modal view also has a direct application to the measurement situation, and agrees with the usual answers given by the environment-induced decoherence approach to classicality ([17], [18]; see also [19]). Consider a quantum measurement as described above, where the reduced states of the measured system $S$ and the measuring apparatus $M$ are

$$\rho_r^S = Tr_{(M)}|\psi\rangle\langle\psi| = \sum_i |\alpha_i|^2 |a_i\rangle\langle a_i| = \sum_i |\alpha_i|^2 \Pi_i^a \qquad (8)$$

$$\rho_r^M = Tr_{(S)}|\psi\rangle\langle\psi| = \sum_i |\alpha_i|^2 |p_i\rangle\langle p_i| = \sum_i |\alpha_i|^2 \Pi_i^p \qquad (9)$$

According to the Vermaas-Dieks interpretation, the preferred context of $S$ is defined by the projectors $\Pi_i^a$ and the preferred context of $M$ is defined by the projectors $\Pi_i^p$. Therefore, also in this case, the observables $A$ of $S$ and $P$ of $M$ acquire actual definite values, whose probabilities are given by the diagonal elements of the diagonalized reduced states.

Although successful in the ideal case, the Kochen-Dieks and the Vermaas-Dieks interpretations, when applied to non-ideal measurements, lead to results that disagree with those obtained in the orthodox collapse interpretation (see [20]-[22]). For instance, if the biorthogonal decomposition is applied to the non-perfectly correlated state $\sum_{ij} \beta_{ij} |a_i\rangle \otimes |p_j\rangle$ (see eq. (4)), the equivalent state $\sum_i \alpha_i' |a_i'\rangle \otimes |p_j'\rangle$ obtained by a change of basis does not select the pointer $P$ as a definite-valued property, but a different observable $P'$ with eigenstates $|p_i'\rangle$ (an analogous argument can be applied to the spectral decomposition case). If the properties ascribed by a modal interpretation are different from those ascribed by collapse, the question is how different they are. In the case of an imperfect measurement, it can be expected that the $\beta_{ij} \neq 0$, with $i \neq j$, be small; then, the difference might be also small. But in the case of a disturbing measurement, the $\beta_{ij} \neq 0$, with $i \neq j$, need not to be small and, as a consequence, the disagreement between the properties ascribed by the modal interpretation and those ascribed by collapse might be unacceptable (see a full discussion in [23]). This fact has been considered by Harvey Brown as a "silver bullet" for killing the modal interpretations (cited in [23]). However, this claim does not apply to the modal-Hamiltonian interpretation, proposed a decade later.

*3.2. The modal-Hamiltonian interpretation*

In most modal interpretations, the preferred context of definite-valued observables depends on the state of the system. This is the case of the traditional Kochen-Dieks interpretation and its Vermaas-Dieks generalization. Jeffrey Bub [24] conceives Bohmian mechanics as the exception, that is, a modal interpretation in which the preferred context does not depend on the state but is defined by an observable of the system, in particular, by the position observable. For this reason, according to Bub, the difficulties that non-ideal measurements pose to the state-depending modal interpretations turn Bohmian mechanics into the only realist non-collapse interpretation still valid. However, this conclusion does not take into account that Bohm's view is not the only reasonable possibility for a modal interpretation with a fixed preferred observable. In fact, the modal-Hamiltonian interpretation ([25]-[29]) endows the Hamiltonian with a determining role, both in the definition of systems and subsystems and in the selection of the preferred context.

By adopting an algebraic perspective, the modal-Hamiltonian interpretation defines a quantum system $S$ as a pair $(\mathcal{O}, H)$ such that (i) $\mathcal{O}$ is a space of self-adjoint operators on a Hilbert space $\mathcal{H}$, representing the observables of the system, (ii) $H \in \mathcal{O}$ is the time-independent Hamiltonian of the system $S$, and (iii) if $\rho_0 \in \mathcal{O}'$ (where $\mathcal{O}'$ is the dual space of $\mathcal{O}$) is the initial state of $S$, it evolves according to the Schrödinger equation. A quantum system so defined can be decomposed in parts in many ways; however, not any decomposition will lead to parts which are, in turn, quantum systems. This will be the case only when the components' behaviors are dynamically independent from each other. In other words, a quantum system can be split into subsystems when there is no interaction among the subsystems:

> **Composite systems postulate:** A quantum system $S: (\mathcal{O}, H)$, with initial state $\rho_0 \in \mathcal{O}'$, is *composite* when it can be partitioned into two quantum systems $S^1: (\mathcal{O}^1, H^1)$ and

$S^2 : (\mathcal{O}^2, H^2)$ such that (i) $\mathcal{O} = \mathcal{O}^1 \otimes \mathcal{O}^2$, and (ii) $H = H^1 \otimes I^2 + I^1 \otimes H^2$, (where $I^1$ and $I^2$ are the identity operators in the corresponding tensor product spaces). In this case, we say that $S^1$ and $S^2$ are *subsystems* of the composite system, $S = S^1 + S^2$. If the system is not composite, it is *elemental*.

With respect to the preferred context, the basic idea of the modal-Hamiltonian interpretation is that the Hamiltonian of the system defines actualization. Any observable that does not have the symmetries of the Hamiltonian cannot acquire a definite actual value, since its actualization would break the symmetry of the system in an arbitrary way:

> **Actualization rule:** Given an elemental quantum system $S : (\mathcal{O}, H)$, the actual-valued observables of $S$ are $H$ and all the observables commuting with $H$ and having, at least, the same symmetries as $H$.

The justification for selecting the Hamiltonian as the preferred observable ultimately lies in the physical relevance of the modal-Hamiltonian interpretation, and in its ability to solve interpretive difficulties. With respect to the first point, the scheme has been applied to several well-known physical situations (free particle with spin, harmonic oscillator, free hydrogen atom, Zeeman effect, fine structure, the Born-Oppenheimer approximation), leading to results consistent with empirical evidence (see [25], Section 5). It is precisely on the basis of these applications that the symmetry requirement for the definite-valued observables can be understood. Let us consider, for instance, the free hydrogen atom, with its quantum numbers: the principal quantum number $n$, the orbital angular momentum quantum number $l$ and the magnetic quantum number $m_l$, which correspond to the eigenvalues of the observables $H$, $L^2$ and $L_z$ respectively. The free hydrogen atom is described in terms of the basis $\{|n, l, m_l\rangle\}$ defined by the CSCO $\{H, L^2, L_z\}$. In this case, the Hamiltonian is degenerate due to its space-rotation invariance. This symmetry of the Hamiltonian makes the selection of $L_z$ a completely arbitrary decision: since space is isotropic, we can choose $L_x$ or $L_y$ to obtain an equally legitimate description of the free atom. The arbitrariness in the selection of the $z$-direction is manifested in spectroscopy by the fact that the spectral lines give no experimental evidence about the values of $L_z$: we have no empirical access to the number $m_l$. The modal-Hamiltonian interpretation agrees with those experimental results since it does not assign a definite value to $L_z$: the actualization of the value of $L_z$ would arbitrarily break the symmetry of the Hamiltonian.

With respect to interpretative issues, the modal-Hamiltonian interpretation was primarily designed to face the problem of contextuality on the basis of an ontology without individuals, according to which quantum systems are bundles of properties, and properties inhabit the realm of possibility, not less real than the domain of actuality (see [30], [31], [32]; for a view that has many points of contact with the modal-Hamiltonian interpretation, see the new transactional interpretation [33], [34], [35]). In spite of this ontological concern, the modal-Hamiltonian interpretation supplies an adequate account of the measurement problem, both in its ideal and its non-ideal versions; this will be the subject of the next section.

## 4. Modal-Hamiltonian interpretation in measurement

Since the actualization rule of the modal-Hamiltonian interpretation is an interpretational postulate that cannot be inferred from the formalism, it has to be assessed by its ability to solve the interpretation difficulties of the theory. Among them, the measurement problem is one of the main challenges.

*4.1. The ideal case*

Let us suppose that we want to obtain the coefficients of the state $|\psi_S\rangle$ of the elemental quantum system $S:(\mathcal{O}_S \subseteq \mathcal{H}_S \otimes \mathcal{H}_S, H_S)$:

$$|\psi_S\rangle = \sum_i \alpha_i |a_i\rangle \qquad (10)$$

where $\{|a_i\rangle\}$ is a basis of $\mathcal{H}_S$, and

$$A = \sum_i a_i |a_i\rangle\langle a_i| \qquad (11)$$

For simplicity, we will assume that the Hamiltonian $H_S$ is non-degenerate:

$$H_S = \sum_i \omega_{Si} |\omega_{Si}\rangle\langle \omega_{Si}| \qquad (12)$$

where $\{|\omega_{Si}\rangle\}$ is also a basis of $\mathcal{H}_S$.

The measuring apparatus is a macroscopic quantum system $M:(\mathcal{O}_M \subseteq \mathcal{H}_M \otimes \mathcal{H}_M, H_M)$; therefore, it has a huge number of microscopic degrees of freedom. As a consequence, the Hamiltonian $H_M$ of $M$, although not completely non-degenerate, will also have a huge number of eigenvalues: the apparatus has an immense number of microscopic possible values $\omega_{Mi}$ of its energy:

$$H_M = \sum_i \omega_{Mi} \Pi_{Mi} \qquad (13)$$

where the set $\{\Pi_{Mi}\}$ of low-dimensional eigenprojectors of $H_M$ spans the Hilbert space $\mathcal{H}_M$ of $M$. The observable $P$ is the pointer observable of the apparatus It must possess different and macroscopically distinguishable eigenvalues $p_i$ in order to play the role of the pointer. This means that it cannot have a number of different eigenvalues as high as that of $H_M$, to the extent that the experimental physicist must be able to discriminate among them (e.g., in the Stern-Gerlach experiment, the pointer has three eigenvalues). In Omnès' terms ([36], [37]), $P$ is a "collective" observable of $M$, that is, a highly degenerate observable that does not "see" the vast majority of the degrees of freedom of $M$:

$$P = \sum_i p_i \Pi_{Pi} \qquad (14)$$

where the set $\{\Pi_{Pi}\}$ of high-dimensional eigenprojectors of $P$ also spans the Hilbert space $\mathcal{H}_M$ of $M$ (see [38]). In measurement, a set $\{|p_i\rangle\}$ of orthogonal eigenvectors will be used to introduce the correlation with the eigenvectors of the measured observable. By construction of the apparatus, the pointer $P$ commutes with $H_M$ for the eigenvectors $|p_i\rangle$ to be stationary and, therefore, the reading of $P$ to be possible.

If $M$ is prepared in a ready-to-measure state $|p_0\rangle$, eigenstate of $P$, the state of the composite system $S+M$ before measurement is given by

$$|\psi_0\rangle = \sum_i \alpha_i |a_i\rangle \otimes |p_0\rangle \qquad (15)$$

In the interaction stage, the systems $S$ and $M$ interact through an interaction Hamiltonian $H_{\text{int}}$. If the measurement is ideal, $H_{\text{int}}$ introduces a perfect correlation between the $|a_i\rangle$ and the $|p_i\rangle$ in such a way that, when the interaction ends, the state $|\psi\rangle$ of $S+M$ is given by eq. (1). In this situation, in spite of the fact that the state $|\psi\rangle$ is a correlated state, $S$ and $M$ turn out to be subsystems of the composite system $S+M$. Therefore, the modal-Hamiltonian actualization rule can be applied to each one of them. Since $[H_M, P] = 0$ and $P$ is much more degenerate (has much more symmetries) than $H_M$, such application results in the fact that both $H_M$ and $P$ are definite-valued observables of $M$.

The $|\alpha_i|^2$ can be obtained by registering the frequencies $fr_i$ of detection of each eigenvalue $p_i$ of $P$ since

$$\Pr(p_i) = \langle p_i | \psi \rangle \langle \psi | p_i \rangle = |\alpha_i|^2 \tag{16}$$

In summary, according to the modal-Hamiltonian interpretation, no matter the value of the measured system's observable, the apparatus' pointer is always definite-valued, and the frequencies of those definite values provide us the correct coefficients of the system's state.

*4.2. The non-ideal case*

As explained above, in the general case the correlation introduced by $H_{int}$ is not perfect; then, when the interaction ends, the state $|\psi\rangle$ of $S + M$ is not given by eq. (1), but by

$$|\psi\rangle = \sum_{ij} \beta_{ij} |a_i\rangle \otimes |p_j\rangle \tag{17}$$

By contrast with other modal interpretations, since $[H_M, P] = 0$ and $P$ has more symmetries than $H_M$, in spite of the imperfection, both $H_M$ and $P$ are definite-valued observables of $M$. The probability of each eigenvalue $p_i$ of $P$ is computed as

$$\Pr(p_i) = \langle p_i | \psi \rangle \langle \psi | p_i \rangle = \sum_n |\beta_{ni}|^2 = |\beta_{ii}|^2 + \sum_{n \neq i} |\beta_{ni}|^2 \tag{18}$$

If the coefficients $\beta_{ni}$, with $n \neq i$, are zero, we are in the ideal measurement case, where $|\beta_{ii}|^2 = |\alpha_i|^2$.

The non-ideal measurement case corresponds to the case in which the coefficients $\beta_{ni}$, with $n \neq i$, are not zero. However, in this case two situations have to be distinguished:

➤ If the $\beta_{ni}$, with $n \neq i$, are small in the sense that $\sum_{n \neq i} |\beta_{ni}|^2 \ll |\beta_{ii}|^2$, then $|\beta_{ii}|^2 \simeq |\alpha_i|^2$. This means that, in the frequency measurement performed by repetition of this single measurement, the coefficients $|\alpha_i|^2$ can be approximately obtained and, therefore, the frequency measurement is *reliable*.

➤ If the $\beta_{ni}$, with $n \neq i$, are not small, then $|\beta_{ii}|^2 \simeq |\alpha_i|^2$ does not hold. Therefore, the result obtained by means of the frequency measurement will be *non reliable*.

Summing up, it is true that, in spite of the fact that ideal measurement is a situation that can never be achieved in practice, physicists usually perform useful measurements. The modal-Hamiltonian account of the measurement process clearly shows that perfect correlation is not a necessary condition for "good" measurements: the coefficients of the system's state at the beginning of the process can be approximately obtained even when the correlation is not perfect, if the reliability condition of small off-diagonal terms is satisfied. Nevertheless, both in the reliable and in the non reliable frequency measurement, in each single measurement *a definite reading of the pointer P is obtained*: the modal-Hamiltonian interpretation is immune to Brown's "silver bullet".

*4.3. The source of non-reliability*

This original modal-Hamiltonian account of non-ideal measurements does not take into account the difference between imperfect measurement and disturbing measurement, because the result of the application of the modal-Hamiltonian actualization rule does not depend on the values of the off-diagonal terms $\beta_{ij}$: the observable $P$, which plays the role of the apparatus' pointer, acquires a definite value in any case. For this reason, in spite of the difference between the two sources of non-ideality, it is sufficient to consider the non-ideal measurement under the following form (see eq. (6)):

$$|\psi\rangle = \sum_{ij} \beta_{ij} |a_i\rangle \otimes |p_j\rangle \qquad (19)$$

In general, the fact that the off-diagonal coefficients be small or not depends on the state of the measured system. All the coefficients $\beta_{ij}$ depend on the initial state of $S$ since they are functions of the $\alpha_i$. This means that the reliability condition $\sum_{n \neq i} |\beta_{ni}|^2 \ll |\beta_{ii}|^2$ does not express a feature of the apparatus itself, but corresponds to the measurement situation as a whole: it can be satisfied for certain initial states and not for others.

These considerations show that, although the reliability condition originally offered by the modal-Hamiltonian interpretation is applicable case by case, it does not provide a condition useful from an experimental viewpoint, since it does not supply a criterion to distinguish between reliable and non reliable measuring apparatus. As we will see in the next section, such a criterion can be obtained when measurement is conceived as an information process, where the measured system is the source of information, the pointer of the measurement apparatus is the destination of information, and the measuring device itself plays the role of the information channel.

**5. An informational account of measurement**

*5.1. Information in Shannon's context*

The seminal work for the mathematical view of information is the paper where Claude Shannon ([39]) introduces a precise formalism designed to solve certain specific technological problems in communication engineering (see also [40]). The most important results obtained by Shannon are related with optimal coding and maximum rate of information transmission. Nevertheless, here we will not be interested in the process of coding, but only in the general situation of information transfer.

From a very general viewpoint, a communication situation can be represented by the following diagram:

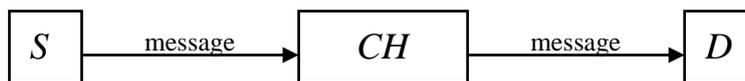

**Figure 1.** Sketch of a communication situation.

where:

– The *source S* generates the message to be received at the destination.
– The *channel CH* is the medium used to transmit the information from source to destination.
– The *destination D* receives the message.

The source $S$ is a system with a range of possible states $s_1,...,s_n$ usually called *letters*, whose respective probabilities of occurrence are $p(s_1),..., p(s_n)$. The amount of information generated at the source by the occurrence of $s_i$ can be defined as

$$I(s_i) = -\log p(s_i) \tag{20}$$

Since *S* produces sequences of states, usually called *messages*, the *entropy of the source S* is computed as a weighted average as

$$H(S) = -\sum_{i=1}^{n} p(s_i) \log p(s_i) \tag{21}$$

Analogously, the destination *D* is a system with a range of possible states $d_1,...,d_m$, with respective probabilities $p(d_1),..., p(d_m)$. The amount of information $I(d_j)$ received at the destination by the occurrence of $d_j$ can be defined as

$$I(d_j) = -\log p(d_j) \tag{22}$$

and the *entropy of the destination D* is computed as

$$H(D) = -\sum_{j=1}^{m} p(d_j) \log p(d_j) \tag{23}$$

In spite of their formal similarity, the entropies of the source $H(S)$ and of the destination $H(D)$ are conceptually different. $H(S)$ is the information produced at the source and, then, only depends on its features (on the distribution of probabilities on its states). $H(D)$, by contrast, is the information that the destination receives and, then, depends not only on the information produced by the source, but also on the channel, that is, on the information lost during transmission and the spurious information not originated in the source. Therefore, the relationship between $H(S)$ and $H(D)$ can be represented in the following diagram (see, e.g., [41], p. 20):

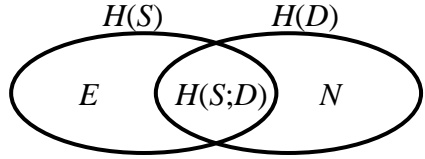

**Figure 2.** Relationship between the entropies of the source and of the destination.

where:

− $H(S;D)$ is the *mutual information*: the average amount of information generated at the source *S* and received at the destination *D*.

− *E* is the *equivocity*: the average amount of information generated at *S* but not received at *D*.

− *N* is the *noise*: the average amount of information received at *D* but not generated at *S*.

As Figure 2 shows, the mutual information can be computed as

$$H(S;D) = H(S) - E = H(D) - N \tag{24}$$

Equivocation *E* and noise *N* are measures of the dependence between the source *S* and the destination *D*:

- If *S* and *D* are completely independent, the values of *E* and *N* are maximum ($E = H(S)$ and $N = H(D)$), and the value of $H(S;D)$ is minimum ($H(S;D) = 0$).

- If the dependence between *S* and *D* is maximum, the values of *E* and *N* are minimum ($E = N = 0$), and the value of $H(S;D)$ is maximum ($H(S;D) = H(S) = H(D)$).

The values of the equivocity $E$ and the noise $N$ are functions not only of the source and the destination, but also of the communication channel $CH$. The introduction of the communication channel leads directly to the possibility of errors in the process of transmission: the channel $CH$ is defined by the matrix $\left[p(d_j/s_i)\right]$, where $p(d_j/s_i)$ is the conditional probability of the occurrence of $d_j$ in the destination $D$ given that $s_i$ occurred in the source $S$, and the elements in any row add up to 1. On this basis, $E$ and $N$ can be computed as

$$E = -\sum_{j=1}^{m} p(d_j) \sum_{i=1}^{n} p(s_i/d_j) \log p(s_i/d_j) \tag{25}$$

$$N = -\sum_{i=1}^{n} p(s_i) \sum_{j=1}^{m} p(d_j/s_i) \log p(d_j/s_i) \tag{26}$$

where $p(s_i/d_j) = p(d_j/s_i)p(s_i)/p(d_j)$. On this basis, the mutual information results

$$H(S;D) = -\sum_{j=1}^{m}\sum_{i=1}^{n} p(s_i,d_j) \log p(s_i,d_j) \tag{27}$$

where $p(s_i,d_j) = p(s_i/d_j)p(d_j) = p(d_j/s_i)p(s_i)$. When the value of $H(S;D)$ is maximum (the values of $E$ and $N$ are minimum, $E = N = 0$), it is said that the channel is *deterministic*: all the information produced at the source arrives at the destination. In turn, in a *noisy channel*, $N \neq 0$, the mutual information is lower than the information of the destination since $H(S;D) = H(D) - N$: the destination receives spurious information that was not produced at the source. In an *equivocal channel*, $E \neq 0$, the mutual information is lower than the information of the source since $H(S;D) = H(S) - E$: part of the information generated at the source gets lost and does not arrive at the destination.

With these elements we have all the tools needed to reconstruct measurement as an informational process.

*5.2. Measurement as an informational process*

Here we will consider a completely general measurement, in which the interaction not only introduces cross terms in the final state, but also disturbs the state of the system. Let us suppose that the calibration of the apparatus is performed by using a known state $|a_k\rangle$ of the measured system $S$, obtained as the result of a filtering-type measurement (see [3], p. 246; for an account of this kind of situations as consecutive measurements in the framework of non-collapse single-system interpretations, such as the modal-Hamiltonian interpretation, see [42], [43]). So, in this case the interaction introduces the following evolution:

$$|a_k\rangle \otimes |p_0\rangle \rightarrow \sum_{ij} \beta_{ij} |a_i^d\rangle \otimes |p_j\rangle \tag{28}$$

where the $\beta_{ij}$ and the $|a_i^d\rangle$ are completely generic; therefore, the $|a_i^d\rangle$ do not need to be orthogonal to each other and may be superpositions of the the $|a_i\rangle$. When this state is expressed in the basis $|a_i\rangle \otimes |p_j\rangle$, it results

$$|a_k\rangle \otimes |p_0\rangle \rightarrow \sum_{ij} \gamma_{ij}^k |a_i\rangle \otimes |p_j\rangle \tag{29}$$

where $\gamma_{ij}^k$ embodies both imperfection and disturbance.

Therefore, if the composite system is prepared in the initial state $|a_k\rangle \otimes |p_0\rangle$, we can measure the frequency $fr_{jk}$ of each value $p_j$ of the pointer $P$. This frequency represents the probability of the fact that the pointer $P$ has the value $p_j$ given that the observable $A$ of the measured system was $a_k$,

$\Pr(p_j/a_k)$. But since the measurement is not ideal, measuring $p_j$ does not guarantee perfect correlation; on the contrary,

$$\Pr(p_j/a_k) = \Pr\left(\left(|a_1\rangle\otimes|p_j\rangle \text{ or } |a_2\rangle\otimes|p_j\rangle \text{ or } ... \text{ or } |a_n\rangle\otimes|p_j\rangle\right)/a_k\right) \tag{30}$$

In turn, since the states $|a_i\rangle\otimes|p_j\rangle$ are orthogonal and, thus, exclusive, this conditional probability can be computed as

$$\Pr(p_j/a_k) = \Pr(|a_1\rangle\otimes|p_j\rangle/a_k) + \Pr(|a_2\rangle\otimes|p_j\rangle/a_k) + ... + \Pr(|a_n\rangle\otimes|p_j\rangle/a_k) \tag{31}$$

and, according to eq. (29),

$$\Pr(p_j/a_k) = |\gamma_{1j}^k|^2 + |\gamma_{2j}^k|^2 + ... + |\gamma_{nj}^k|^2 = \sum_i |\gamma_{ij}^k|^2 \tag{32}$$

Once calibration has been performed, a generic measurement can easily be characterized. When the unknown state of the measured system $S$ is a superposition, the evolution induced by the interaction is (see eq. (29))

$$\sum_k \alpha_k |a_k\rangle \otimes |p_0\rangle \;\rightarrow\; \sum_k \alpha_k \sum_{ij} \gamma_{ij}^k |a_i\rangle\otimes|p_j\rangle = \sum_{ij}\sum_k \alpha_k \gamma_{ij}^k |a_i\rangle\otimes|p_j\rangle \tag{33}$$

If this eq. (33) is compared with eq. (6), it is easy to recognize the dependence of the $\beta_{ij}$ with respect to the coefficients $\alpha_k$ of the initial state:

$$\beta_{ij} = \sum_k \alpha_k \gamma_{ij}^k \tag{34}$$

In this situation, we measure the frequency $fr_j$ of each value $p_j$ of the pointer $P$, which represents the probability $\Pr(p_j)$ of the value $p_j$, given by

$$\Pr(p_j) = \sum_i |\beta_{ij}|^2 = \sum_i \left|\sum_k \alpha_k \gamma_{ij}^k\right|^2 \tag{35}$$

If the measurement had been *perfect* (perfect correlation) but *disturbing*, then $\gamma_{ij}^k = \tilde{\gamma}_{ij}^k \delta_{ij}$. If the measurement had been *non-disturbing* but *imperfect*, then $\gamma_{ij}^k = \tilde{\gamma}_{ij}^k \delta_{ik}$. If the measurement had been *ideal*, then $\gamma_{ij}^k = \delta_{ji}\delta_{ik}$ and, as a consequence,

$$\Pr(p_j) = \sum_i \left|\sum_k \alpha_k \delta_{ji}\delta_{ik}\right|^2 = |\alpha_j|^2 = \Pr(a_j) \tag{36}$$

But in the generic case considered here there are both imperfection and disturbance and, thus, the measured frequencies do not give direct information about the coefficients of the measured state. Moreover, the original reliability condition $\sum_{n\neq i}|\beta_{ni}|^2 \ll |\beta_{ii}|^2$ cannot be used to know if the measuring apparatus is reliable or not, because the coefficients $\beta_{ij}$ are functions of the coefficients $\alpha_k$ of the unknown initial state. In this case, the informational conception of the quantum measurement proves to be fruitful.

If the just described generic measurement is conceived as a process of transference of information, the measured quantum state is the source $S$ of information, with a range of possible "states" $a_1,...,a_n$, that is, the eigenvalues of the observable $A$, whose respective probabilities of occurrence are $\Pr(a_j) = |\alpha_j|^2$. In turn, the pointer $P$ is the destination, with a range of possible "states" $p_1,...,p_n$, that is, the eigenvalues of $P$ with probabilities of occurrence $\Pr(p_j)$, respectively. On the other hand, the measuring apparatus is the channel, which can be mathematically characterized independently of the peculiar features of source and destination. In particular, the channel is defined by the matrix

$\left[\Pr\left(p_j/a_k\right)\right]$ which, as explained above, can be empirically determined prior to any particular measurement, in the process of calibration.

From this informational perspective, we can say that:

➢ When the channel is *deterministic*, then the measurement is *ideal*. In this case, all and only the information of the measured quantum state is recovered by the pointer. The perfect correlation introduced by the measuring device is embodied in the fact that the conditional probabilities that define the channel are trivial: $\Pr\left(p_j/a_k\right) = \delta_{jk}$.

➢ When the channel is *equivocal or/and noisy*, then the measurement is *non-ideal*. In this case, there is some loss of relevant information $E$ or/and addition of spurious information $N$ through the process: this is embodied in the fact that the conditional probabilities that define the channel are not trivial. But since those probabilities characterize the measuring device through calibration, they can be used to give a criterion of reliability independent of the particular measurement carried out:

– If the conditional probabilities are approximately trivial, $\Pr\left(p_j/a_k\right) \simeq \delta_{jk}$, then $\Pr\left(p_j\right) \simeq \Pr\left(a_j\right) = |\alpha_k|^2$. This means that, in the frequency measurements performed by repetition of the single measurement, the measured frequencies $fr_j$ approximately supply the value of the coefficients $|\alpha_i|^2$ and, therefore, the frequency measurement is *reliable*.

– If the conditional probabilities are not approximately trivial, then $\Pr\left(p_j\right) \simeq \Pr\left(a_j\right) = |\alpha_k|^2$ does not hold. Therefore, the results obtained by means of frequency measurements will be *non reliable*.

Besides offering a reliability criterion that depends on the previously calibrated measuring apparatus and not on the state to be measured, the informational view of measurement supplies the tools to quantify the degree of non-reliability of a given device. In fact, given the conditional probabilities $\Pr\left(p_j/a_k\right)$ of the channel-apparatus, mutual information, equivocity and noise can be computed. Of course, these magnitudes will depend on the particular quantum state to be measured. Nevertheless, the reliability of different measuring devices can be compared on the basis of some standard initial state. A reasonable standard is the situation in which the information generated at the source is maximum, that is, all its states are equiprobable. In this case, the value of the $n$ coefficients $\alpha_k$ of the initial quantum state of the system $S$ will be $\sqrt{1/n}$; so $\Pr(s_i) = 1/n$. If the frequencies $fr_j$ of detection of the eigenvalues $p_j$ of $P$, which represent the probabilities $\Pr\left(p_j\right)$ of the values $p_j$, are measured in this situation, then $E$, $N$ and $H(S;D)$ can be computed with the eqs. (25), (26) and (27). On this basis, the degree of reliability $R$ of the measuring device can be defined as:

$$R = \frac{H(S;D)}{H(S;D)_{\max}} = \frac{H(S;D)}{H(S)} = -\frac{H(S;D)}{\log n} \tag{37}$$

where $H(S;D)_{\max} = H(S) = -\log n$ corresponds to the ideal case. Therefore, in the ideal measurement, $R = 1$.

Moreover, in the case of non-reliability, the values of the equivocity $E$ and the noise $N$ computed in this situation give clues about the source of this undesirable feature. Let us recall that in a noisy channel, $N \neq 0$, the destination receives spurious information that was not produced at the source, and in an equivocal channel, $E \neq 0$, part of the information generated at the source does not arrive at the destination. If we want that every state of the destination lets us know which state of source occurred, it is necessary that the backward probabilities $p(s_i/d_j)$ have the value 0 or 1, and this happens in an equivocation-free channel, $E \neq 0$, no matter the value of noise. This explains why noise does not prevent observation: indeed, practical situations of communication usually include noisy channels, and much technological effort is devoted to design appropriate filters or shielding to block the noise bearing spurious signal. On the contrary, an equivocal channel leads to an information loss that cannot be remediated by means of filters, but requires the addition of redundant information. In other words, in communication the different sources of imperfection are not equally serious, and demand different strategies for their solution. Of course, communication is different than measurement: whereas in communication the states of the source are previously known and can be controlled at will, in measurement the state of the measured system is unknown. Nevertheless, when the performances of different measuring apparatuses are compared on the basis of a standard known state, knowing whether the source of non-reliability is equivocity or noise (or both) may give hints about what kind of actions on the apparatus may be more efficient to solve the problem.

## 6. Conclusions

In this paper we have considered the account of quantum measurement supplied by the modal-Hamiltonian interpretation, which explains the definite-valuedness of the pointer reading both in ideal and non-ideal measurements, and distinguishes between reliable and non-reliable situations in the non-ideal case. But since the reliability condition originally proposed depended on the state to be measured, in this paper we improved that proposal by offering a reliability condition that characterizes the measuring apparatus in itself. To fulfill this task, we reconstructed the measurement process as an informational situation, where the measured state is the source of information, the measuring pointer is the destination of information, and the measuring device plays the role of the information channel. Furthermore, this reconstruction allowed us to quantify reliability and to characterize the distinction between different sources of imperfection.

On the basis of this analysis of the quantum measurement, now we can explicitly formulate the conditions that a quantum process must satisfy to be considered a single measurement by the modal-Hamiltonian interpretation:

(a) There must be two quantum systems: a system to be measured, $S:(\mathcal{O}, H_S)$, and a measuring apparatus, $M:(\mathcal{O}, H_M)$.

(b) The apparatus $M$ must be constructed in such a way to have a pointer observable $P$ such that (i) $[H_M, P] = 0$, (ii) $P$ has, at least, the same degeneracy as $H_M$, and (iii) the eigenvalues of $P$ are different and macroscopically distinguishable. As argued, these conditions are physically reasonable independently of this interpretation.

(c) During a certain period, $S$ and $M$ interact through an interaction Hamiltonian $H_{int}$ intended to introduce a correlation between an observable $A$ of $S$ and the pointer $P$ of $M$.

Let us notice that the requirement of perfect correlation is not included as a defining condition of single measurement: even if the correlation is not perfect, the pointer $P$ always acquires a definite value in each single measurement. Nevertheless, the frequency measurement resulting from the repetition of single measurements is not always *reliable*, and the reliability condition can be precisely defined.

Two final interpretive remarks. First, the informational account of quantum measurement just proposed is not univocally tied to the modal-Hamiltonian interpretation: it might be applied to a different interpretation. However, this would be possible provided that the considered interpretation gives an adequate explanation of the definite reading of the apparatus' pointer in any case, independently of the degree of correlation established in the interaction. Second, this informational view of quantum measurement is independent of the interpretation of information adopted. In particular, information can be conceived as an epistemic feature or as a physical item ([44]-[46]), or can be deprived of physical meaning ([47]-[48]): the interpretive stance regarding information does not affect the reconstruction of measurement as a process of transfer of information.

**Acknowledgments**

We are grateful to the participants of the workshop *What is quantum information?* (Buenos Aires, May of 2015), Jeffrey Bub, Adán Cabello, Dennis Dieks, Armond Duwell, Christopher Fuchs, Robert Spekkens and Christopher Timpson, for the stimulating and lively discussions about the concept of information. This paper was partially supported by a Large Grant of the Foundational Questions Institute (FQXi), and by a grant of the National Council of Scientific and Technological Research (CONICET) of Argentina.